\documentstyle[12pt,epsfig,here]{article}
\begin{document}
\newcommand{\beq}{\begin{equation}}
\newcommand{\eeq}{\end{equation}}
\def\beqn{\begin{eqnarray}}
\def\eeqn{\end{eqnarray}}

\newcommand{\Tr}{{\rm Tr}\,}
\newcommand{\E}{{\cal E}}

\newcommand{\ntwo}{${\cal N}=2\;$}
\newcommand{\none}{${\cal N}=1\;$}
\newcommand{\noneh}{${\cal N}=\,
^{\mbox{\small 1}}\!/\mbox{\small 2}\;$}
\newcommand{\vp}{\varphi}
\newcommand{\ve}{\varepsilon}
\newcommand{\pt}{\partial}

\newcommand \vev [1] {\langle{#1}\rangle}
\newcommand\lr[1]{{\left({#1}\right)}}

\begin{flushright}
FTPI-MINN-08/15, UMN-TH-2646/08
\end{flushright}
 
 \vspace{1cm}
 
 \begin{center}
 {\bf {\Large\bf Theoretical Developments in SUSY}}
 
  \vspace{1cm}
 
{\large M. Shifman}

 \vspace{8mm}
 
{\it  William I. Fine Theoretical Physics Institute,
University of Minnesota,
Minneapolis, MN 55455}

 \vspace{3cm}
 
 {\em Abstract}

\end{center}

\vspace{2mm}

Invited Talk at PLANCK 2008, Barcelona, Spain, 	
19 -- 23 May, 2008

 \vspace{3mm}
 
\underline{Main topics:}

\vspace{3mm}

$\bullet$ Heterotic strings from ${\mathcal N}=1$ gauge theories;
 \vspace{1mm}

$\bullet$ Planar equivalence and emergent center symmetry in QCD-like theories; 
 \vspace{1mm}

$\bullet$ Exact result for gluon scattering amplitudes in ${\mathcal N}=4$
(dual conformality).

\newpage

\section{Non-Abelian heterotic strings in ${\mathcal N}=1$: setting the stage}

Seiberg and Witten presented \cite{SW} the first ever demonstration of the dual Meissner effect
in non-Abelian theory, a celebrated analytic proof of linear
confinement, which caused much excitement in the community. The Seiberg--Witten 
flux tubes are essentially Abelian (of the Abrikosov--Nielsen--Olesen 
type), so that the hadrons they create are not alike
those in QCD \cite{HSZ}. 

What do we mean when we speak of Abelian versus non-Abelian flux tubes?
In the former case, gauge dynamics relevant to distances where
the tubes are formed is that of an Abelian theory (although short-distance dynamics
can well be non-Abelian, as in the Seiberg--Witten case). In the latter case,
in the infrared, at distances relevant to the tube formation, dynamics is determined
by non-Abelian theory, with all gauge bosons equally operative. 
Correspondingly, we can speak of Abelian versus non-Abelian confinement.
There are reasons to believe that
no phase transition occurs between these two regimes in the Seiberg--Witten 
solution.\footnote{It was argued \cite{Shifman:2008ja} that, under certain conditions,
transition from Abelian to non-Abelian confinement is smooth
in non-supersymmetric QCD compactified on $S_1\times R_3$.}
However, in the limit of large-$\mu$ deformations, when a non-Abelian regime presumably sets
in and non-Abelian strings develop in the model considered by Seiberg and Witten, theoretical control is completely lost. What was badly needed and sought for was a model
in which non-Abelian strings develop in a fully controllable manner, i.e. at weak coupling.

Ever since, searches for non-Abelian
flux tubes and non-Abelian mono\-poles continued,
with a decisive breakthrough in 2003-04 \cite{HT,auzzietal}. By that time the program of
finding field-theoretical analogs of all basic constructions
of string/D-brane theory was in full swing.
BPS domain walls, analogs of D branes, had been identified in
supersymmetric Yang--Mills theory \cite{DS}. It had been demonstrated that
such walls support gauge fields localized on them. BPS saturated 
string-wall junctions had been constructed \cite{SY1}.

\subsection{Non-Abelian flux tubes and monopoles}
\label{naft}

\begin{figure}
\begin{center}
\psfig{figure=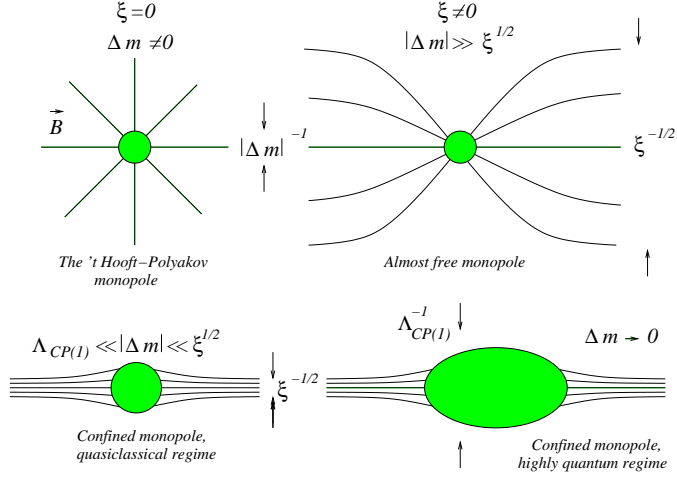,height=2.5in}
\end{center}
\caption{\small Various regimes for monopoles and strings.}
\label{fig:radish}
\end{figure}

Non-Abelian strings were first found 
in ${\mathcal N}=2$  super-Yang--Mills theories with U(2)$_{\rm gauge}$ and two matter hypermultiplets \cite{HT,auzzietal}. 
The \ntwo vector multiplet
consists of the  U(1)
gauge field $A_{\mu}$ and the SU(2)  gauge field $A^a_{\mu}$,
(here $a=1,2,3$), and their Weyl fermion superpartners
($\lambda^{1}$, $\lambda^{2}$) and
($\lambda^{1a}$, $\lambda^{2a}$), plus
complex scalar fields $a$, and $a^a$.  The  global SU(2)$_R$ symmetry inherent to
 \ntwo   models manifests itself through rotations
$\lambda^1 \leftrightarrow \lambda^2$.

The quark multiplets consist
of   the complex scalar fields
$q^{kA}$ and $\tilde{q}_{Ak}$ (squarks) and
the  Weyl fermions $\psi^{kA}$ and
$\tilde{\psi}_{Ak}$, all in the fundamental representation of 
the SU(2) gauge group
($k=1,2$ is the color index
while $A$ is the flavor index, $A=1,2$).
The scalars $q^{kA}$ and ${\bar{\tilde q}}^{\, kA}$
form a doublet under the action of the global
SU(2)$_R$ group.  The quarks and squarks have a U(1) charge too.

If one introduces a non-vanishing Fayet--Iliopoulos parameter $\xi$
the theory develops isolated quark vacua,
in which the gauge symmetry is fully Higgsed, and all elementary excitations are massive.
In the general case, two matter
mass terms allowed by  ${\mathcal N}=2$
are unequal, $m_1\neq m_2$. 
There are free parameters whose interplay
determines dynamics of the theory:
the Fayet--Iliopoulos parameter $\xi$, the mass difference
$\Delta m$ and a dynamical scale parameter
$\Lambda$, an analog of the QCD scale $\Lambda_{\rm QCD}$ (Fig.~\ref{fig:radish}). 
Extended supersymmetry guarantees that some crucial dependences are holomorphic,
and there is no phase transition.

Both the gauge and flavor symmetries of the model are
broken by the squark condensation. 
All gauge bosons acquire the same masses (which are of the order of 
inverse string thickness). A global diagonal
combination of color and flavor groups, SU$(2)_{C+F}$, survives the breaking
(the subscript $C+F$ means a combination of global color and flavor groups). 

While SU$(2)_{C+F}$ is
the symmetry of the vacuum, the flux tube solutions break it spontaneously.
This gives rise to orientational moduli on the string world sheet. 

The bulk theory is characterized by three parameters of dimension of mass:
$\xi$, $\Delta m$, and $\Lambda$.
As various parameters vary, the theory under consideration evolves in a very
graphic way, see Fig.~\ref{fig:radish}. At $\xi=0$ but 
$\Delta m \neq 0$
(and $\Delta m \gg \Lambda$) it presents a very clear-cut example
of a model with the standard 't Hooft--Polyakov monopole.
The monopole is unconfined  --- the flux tubes are not yet formed.

Switching on $\xi\neq 0$ traps the magnetic fields inside
the flux tubes, which are weak as long as $\xi\ll\Delta m$.
The flux tubes change the shape of the monopole far away from its core,
leaving the core essentially intact. Orientation of the
chromomagnetic field inside the flux tube is essentially fixed. 
The flux tubes are Abelian.

With $|\Delta m|$ decreasing, 
fluctuations in the orientation of the 
chromomagnetic field inside the flux tubes grow. 
Simultaneously, the monopole 
which no loner resembles the 't Hooft--Polyakov monopole, is 
seen as a string junction.

Finally, in the limit $\Delta m\to 0$
the transformation is complete. A global SU(2) symmetry restores
in the bulk. Orientational moduli
develop on the string worldsheet making it non-Abelian.
The string worldsheet theory is CP(1) (CP$(N-1)$ for 
generic values of $N$). 
Two-dimensional CP$(N-1)$ models with four supercharges
are asymptotically free. They have $N$ distinct vacuum states.

Each vacuum state of the worldsheet CP$(N-1)$ theory
presents a distinct string from the standpoint of the bulk theory.
There are $N$ species of such strings; they have degenerate 
tensions $T_{\rm st} =2\pi\xi$.
(The ANO string tension is $N$ times larger.)

\begin{figure}
\begin{center}
\psfig{figure=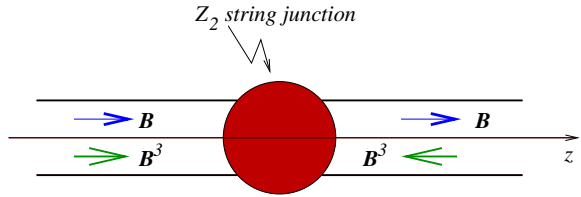,height=1in}
\end{center}
\caption{\small $Z_2$  string junction.}
\label{z2sj}
\end{figure}

Two different strings can form a stable junction.
Figure~\ref{z2sj} shows this junction in the limit
\beq
\Lambda_{{\rm CP}(1)}\ll |\Delta m| \ll \sqrt{\xi}
\label{limi}
\eeq
corresponding to the lower left corner in Fig.~\ref{fig:radish}.
The magnetic fluxes of the U(1) and SU(2) gauge groups
are oriented along the $z$ axis. In the limit (\ref{limi})
the SU(2) flux is oriented along the third axis in the internal space.
However, as $|\Delta m| $ decreases, fluctuations
of $B_z^a$  in the internal space grow, and at $\Delta m\to 0$
it has no particular orientation in SU(2) (the lower right corner of Fig.~\ref{fig:radish}).
In the language of the worldsheet theory this phenomenon
is due to restoration of the O(3) symmetry in the quantum vacuum of
the CP(1) model. 

The junctions of degenerate strings present what remains of the
monopoles in this highly quantum regime \cite{SY2,HT2}. It is remarkable that,
despite the fact we are deep inside the highly quantum regime,
holomorphy allows one to exactly calculate the mass of
these monopoles. This mass is given by the expectation value of the
kink central charge in the worldsheet CP$(N-1)$ model
(including the anomaly term), $M_M \sim N^{-1}\, \langle R \,  \psi_L^\dagger\,\psi_R\rangle$. 
 
 \subsection{Towards ${\mathcal N}=1$}
 
The unwanted feature of ${\mathcal N}=2$ theory,
making it less similar to QCD, is the presence of the adjoint chiral
superfields ${\mathcal A}$ and ${\mathcal A}^a$. One can get rid of them making them heavy. 
To this end we can endow the adjoint superfield
with a mass term of the type
$\mu {\mathcal A}^2$. More exactly, we will consider the ${\mathcal N}=1$
preserving deformation superpotential
\beq
{\mathcal W} = \frac{\mu}{2}\left[ {\mathcal A}^2 + \left({\mathcal A}^a\right)^2\right]
\label{mcsp}
\eeq
where ${\mathcal A}$ and ${\mathcal A}^a$ are adjoint chiral superfields.
Now, supersymmetry of the bulk model becomes ${\mathcal N}=1$.
At large $\mu$ the adjoint fields decouple.

With the deformation  superpotential (\ref{mcsp})
the 1/2 BPS classical flux tube
solution stays the same as in the absence of this superpotential \cite{Shifman:2005st}. 
Moreover, the number of the boson and fermion zero modes,
which become moduli fields on the string worldsheet,
does not change either. 
For the fermion zero modes this statement follows from an index theorem proved
in \cite{GSY}. If the string solution and the number of zero modes remain the
same, what can one say about the string worldsheet theory?

\section{Worldsheet theory on strings in ${\mathcal N}=1$ bulk theories}
\label{wtos}

The discovery of non-Abelian strings in
${\mathcal N}=1$ bulk theories \cite{Shifman:2005st} was a crucial step on the way to
the desired ${\mathcal N}=0$ theories. It turns out that these strings are
quite remarkable. One can call them heterotic non-Abelian strings:
the corresponding worldsheet theory is a chiral
${\mathcal N}=(0,2)$ extension of the 
bosonic CP$(N-1)$ model which was unknown previously!

${\cal N}=2$ SUSY Yang--Mills theories which support
non-Abelian flux tubes have eight supercharges. 
The flux tube solutions are 1/2 BPS-saturated. Hence, 
the effective low-energy theory of the moduli fields
on the string worldsheet must have four supercharges. 
The bosonic moduli consist of two groups: two translational moduli $\left(x_0\right)_{1,2}$
corresponding to translations in the plane perpendicular to the string axis,
and two orientational moduli whose interaction is described by CP(1).
The fermion moduli also split in two groups:
four supertranslational moduli $\zeta_L ,\, \zeta_L^\dagger ,\,\zeta_R ,\, \zeta_R^\dagger$ 
plus four superorientational moduli.
${\cal N}=2$ supersymmetry in the bulk and on the worldsheet guarantees 
that $\left(x_0\right)_{1,2}$ and $\zeta_{L,R}$ form a free field theory on the worldsheet
completely
decoupling from (super)orientational moduli, which in turn form 
${\mathcal N}=(2,2)$ supersymmetric CP(1) model. 

\subsection{Heterotic models on the worldsheet}
\label{41}

What happens when one deforms the bulk theory
to break ${\cal N}=2$ down to ${\cal N}=1$? Now we have four supercharges in the
bulk and expect two supercharges on the worldsheet.
If supertranslational sector continued to be decoupled from the superorientational one
(which seemed to be a reasonable assumption)
supersymmetrization of the orientational and
translational modes would occur separately. It is well known
that the requirement of two supercharges in CP(1)
automatically leads to a nonchiral model
with extended supersymmetry, ${\cal N}=(2,2)$,
with four supercharges (for a review see e.g. \cite{Novikov:1984ac}).
This was the line of reasoning Yung and I followed
in 2005  \cite{Shifman:2005st} in arguing that non-Abelian strings obtained
in the 
${\cal N}=1$ bulk theories 
have an ``accidentally" enhanced supersymmetry. As we will see
shortly, the assumed  decoupling does not take place.

Edalati and Tong noted \cite{ET} that, with two supercharges
on the worldsheet, only  $\zeta_L,\,\,\zeta_L^\dagger $
remain protected. At the same time, $\zeta_R$ can and does mix
with the superorientational moduli. Edalati and Tong outlined
a general structure of the chiral ${\mathcal N}= (0,2)$ generalization
of CP(1). Derivation of the heterotic CP(1) model from the 
bulk theory was carried out in Ref.~\cite{shyu}.
In this model the right- and left-moving fermions acquire different interactions; hence,
the flux tube becomes heterotic!

The Lagrangian of the heterotic CP(1) model is
\beqn
&&
L_{{\rm heterotic}}= 
\zeta_R^\dagger \, i\partial_L \, \zeta_R  + 
\left[\gamma \, \zeta_R  \,R\,  \big( i\,\partial_{L}\phi^{\dagger} \big)\psi_R
+{\rm H.c.}\right] -g_0^2|\gamma |^2 \left(\zeta_R^\dagger\, \zeta_R
\right)\left(R\,  \psi_L^\dagger\psi_L\right)
\nonumber
\\[4mm]
&&
+
G\, \left\{\rule{0mm}{5mm}
\partial_\mu \phi^\dagger\, \partial^\mu\phi  
+\frac{i}{2}\big(\psi_L^\dagger\!\stackrel{\leftrightarrow}{\partial_R}\!\psi_L 
+ \psi_R^\dagger\!\stackrel{\leftrightarrow}{\partial_L}\!\psi_R
\big)\right.
\nonumber
\\[4mm] 
&&
-\frac{i}{\chi}\,  \big[\psi_L^\dagger \psi_L
\big(\phi^\dagger \!\stackrel{\leftrightarrow}{\partial_R}\!\phi
\big)+ \psi_R^\dagger\, \psi_R
\big(\phi^\dagger\!\stackrel{\leftrightarrow}{\partial_L}\!\phi
\big)
\big]
-
\frac{2(1- g_0^2 |\gamma |^2)}{\chi^2}\,\psi_L^\dagger\,\psi_L \,\psi_R^\dagger\,\psi_R
\Big\}\,,
\label{AAone}
\eeqn
where $G$ is the Fubini--Study metric, 
\beq
G = \frac{2}{g_0^2}\, \frac{1}{\left(1 +|\phi|^2\right)^2}\,,
\eeq
$R$ stands for the Ricci tensor, and
\beq
\partial_L =\frac{\partial}{\partial t} +\frac{\partial}{\partial z}\,,\qquad
\partial_R =\frac{\partial}{\partial t} - \frac{\partial}{\partial z}\,.
\eeq
The constant $\gamma$ in Eq. (\ref{AAone}) 
is the parameter which determines the ``strength" of the heterotic deformation,
and the left-right asymmetry in the fermion sector. It is related to
the parameter $\mu$ in Eq.~(\ref{mcsp}) as follows:
\beq
\gamma  = \frac{1}{g_0}\,\frac{\delta }{\sqrt{1+2|\delta |^2}}
\eeq
where $\delta$ is known in two limits \cite{shyu},
\beq
\delta = \left\{
\begin{array}{l}
{\rm const.}\, \mu\,,\quad \mu\,\,\, {\rm small},  \\[3mm]
{\rm const.} \, \sqrt{\rule{0mm}{4mm}\ln\mu} \,,\quad \mu\,\,\, {\rm large}.
\end{array}
\right.
\eeq
The second and third lines in Eq.~(\ref{AAone}) are the same as in the
conventional ${\mathcal N}=(2,2)$ CP(1) model, except the
last coefficient.

Generalization for arbitrary $N$
(i.e. the ${\mathcal N}=(0,2)$ deformed CP$(N-1)$ model) is as follows \cite{shyu}:
\beqn
L_{{\rm heterotic}} && 
= 
\zeta_R^\dagger \, i\partial_L \, \zeta_R  + 
\left[\gamma\, g_0^2 \, \zeta_R  \, G_{i\bar j}\,  \big( i\,\partial_{L}\phi^{\dagger\,\bar j} \big)\psi_R^i
+{\rm H.c.}\right]
\nonumber
\\[4mm]
&&
 -g_0^4\, |\gamma |^2 \,\left(\zeta_R^\dagger\, \zeta_R
\right)\left(G_{i\bar j}\,  \psi_L^{\dagger\,\bar j}\psi_L^i\right)
\nonumber
\\[4mm]
&&
+G_{i\bar j} \big[\partial_\mu \phi^{\dagger\,\bar j}\, \partial_\mu\phi^{i}
+i\bar \psi^{\bar j} \gamma^{\mu} D_{\mu}\psi^{i}\big]
\nonumber
\\[4mm]
&&
- \frac{g_0^2}{2}\left( G_{i\bar j}\psi^{\dagger\, \bar j}_R\, \psi^{ i}_R\right)
\left( G_{k\bar m}\psi^{\dagger\, \bar m}_L\, \psi^{ k}_L\right)
\nonumber
\\[4mm]
&&
+\frac{g_0^2}{2}\left(1-2g^2_0|\gamma|^2\right)
\left( G_{i\bar j}\psi^{\dagger\, \bar j}_R\, \psi^{ i}_L\right)
\left( G_{k\bar m}\psi^{\dagger\, \bar m}_L\, \psi^{ k}_R\right).
\label{cpn-1g}
\eeqn

Introduction of a seemingly rather insignificant heterotic deformation
drastically changes dynamics of the CP(1) model, leading to spontaneous SUSY breaking. 
This is rather obvious at small $\mu$. Indeed, 
the supercurrent of the deformed
model acquires  extra terms proportional to
$\gamma \left\{R \,  \psi_R^\dagger\,\psi_L\right\} \zeta_R^\dagger$
at small $\mu$. In this limit the expression in the braces can be evaluated in the undeformed
CP(1) model. As well known (see e.g. \cite{Novikov:1984ac}),
a nonvanishing bifermion condensate $\langle R \,  \psi_R^\dagger\,\psi_L\rangle 
\sim \pm \Lambda$ develops in this model ($\Lambda$ is the scale parameter)
labeling two distinct vacua. Thus, the additional terms in the supercurrent
emerging in the deformed theory (at small $\gamma$)
have the form
\beq
\Delta J_{\rm sc} =\gamma \, \langle R \,  \psi_R^\dagger\,\psi_L\rangle \,
\zeta_R^\dagger\,,\quad \Delta J_{\rm sc}^\dagger = \gamma \, \langle R \,  \psi_L^\dagger\,\psi_R\rangle \,
\zeta_R\,.
\label{golz}
\eeq
Since $\zeta_R$ is strictly massless, Eq.~(\ref{golz})
clearly demonstrates that $\zeta_R$ is a Goldstino, with the residue $\langle R \,  \psi_R^\dagger\,\psi_L\rangle$. Supersymmetry is spontaneously broken, with the
vacuum energy 
\beq
{\mathcal E}_{\rm vac} = |\gamma|^2\,
\left|\langle R \,  \psi_R^\dagger\,\psi_L\rangle
\right|^2
\eeq
times a numerical factor, one and the same for both vacua. A nonvanishing ${\mathcal E}_{\rm vac}$ for arbitrary values of $\gamma$ in heterotically deformed $CP(N-1)$ models was obtained in \cite{shyu2}
using large $N$ expansion.  The very possibility of the spontaneous supersymmetry breaking
is due to the fact that Witten's index $I_W$ of the
deformed theory vanishes, in sharp contradistinction with the 
undeformed conventional
${\mathcal N}=(2,2)$ model where $I_W =N$.
Spontaneous breaking of SUSY in heterotic $CP(N-1)$
was anticipated in \cite{Tong:2007qj}. 

\subsection{Large-$N$ solution of the heterotic CP$(N-1)$ models}

Solution of the heterotic model (\ref{cpn-1g}) in the large-$N$
limit was found in~\cite{shyu2} basing on the pattern
suggested by Witten long ago \cite{wln}. Here I will briefly describe a general structure of the solution which depends on a single scaling variable
\beq
u = \frac{16\pi}{N}\,\frac{1}{g_0^2}\,|\delta |^2\,,
\eeq
where the deformation parameter $\delta$ was introduced in Sect.~\ref{41} while
$g_0^2$ is the coupling of the CP$(N-1)$ model related to the bulk gauge coupling
as $g_0^{-2} =2\pi\, g_2^{-2}$, see e. g. the review paper \cite{syr}. 

For arbitrary values of the deformation parameter
the chiral condensate takes the form
\beq
\langle R \,  \psi_R^\dagger\,\psi_L\rangle = \Lambda\, \exp
\left(-\frac{u}{2} +\frac{2\pi\, i k}{N}
\right)
\eeq
where the integer $k$ ($k = 1,2, ...,N$) labels $N$ distinct degenerate vacua 
of the theory  corresponding to the
spontaneous breaking of an axial $Z_N$ symmetry. At large $u$ the above order parameter becomes small. The ${\mathcal N}= (0,2)$
supersymmetry is always spontaneously broken as long as $u\neq 0$,
\beq
{\mathcal E}_{\rm vac}=\frac{N}{4\pi}\,\Lambda^2\,\left(1-e^{-u}\right)
\sim \left\{
\begin{array}{l}
u\,N^{-1}\,\left|\langle R \,  \psi_R^\dagger\,\psi_L\rangle
\right|^2\,,\quad u\to 0\\[4mm]
\frac{N}{4\pi}\,\Lambda^2\,, \quad u\to\infty\,.
\end{array}
\right.
\eeq
The theory has a massless Goldstino.
At small $u$ its role is played by $\zeta_R$,
while in the large-$u$ limit $\psi_R$ becomes massless.
In the large $u$ limit, when $\langle R \,  \psi_R^\dagger\,\psi_L\rangle$ is small,
the low-energy effective theory contains, apart from bosonic states,
a single fermion: the massless Goldstino.

From the ${\mathcal N}=1$ bulk theory standpoint, the spontaneous breaking
of the ${\mathcal N}= (0,2)$
supersymmetry on the worldsheet means
the loss of BPS saturation of the heterotic strings 
 due to nonperturbative effects.

\section{Planar Equivalence}
\label{pleq}

Planar equivalence is equivalence in the large-$N$ limit of
distinct QCD-like theories in their common sectors (see \cite{asv}).
Most attention received equivalence between SUSY gluodynamics
and its orientifold  and $Z_2$ orbifold  daughters.
The Lagrangian of the parent theory is
\beq
{\mathcal L}= -\frac{1}{4g_P^2} \, G_{\mu\nu}^a G_{\mu\nu}^a
+\frac{i}{g_P^2}\, \lambda^{a\alpha} D_{\alpha\dot\beta}\bar\lambda^{a\dot\beta}
\eeq
where $\lambda^{a\alpha}$ is the gluino (Weyl) field in the adjoint
representation of SU$(N)$, and $g_P^2$ stands for the coupling constant in the parent theory. The orientifold daughter is obtained by
replacing $\lambda^{a\alpha}$ by the {\em Dirac} spinor in the two-index
(symmetric or antisymmetric) representation (to be referred to as orienti-S
or orienti-AS). The gauge coupling stays intact. 
To obtain the $Z_2$ orbifold daughter (to be referred to as orbi) we must pass
to the gauge group SU$(N/2)\times$SU$(N/2)$, replace $\lambda^{a\alpha}$
by a bifundamental Dirac spinor, and rescale 
the gauge coupling, $g_D^2=2g_P^2$. 

\subsection{Brief history}

 Genesis of planar equivalence can be traced
to string theory. In 1998
Kachru and Silverstein studied  \cite{Kachru:1998ys}
various orbifolds of $R^6$ within the AdS/CFT correspondence, of which I will speak later. Starting from ${\cal N}=4$, they obtained distinct --- but equivalent in the infinite-$N$ limit  ---
four-dimensional daughter gauge field theories with matter, with varying degree of supersymmetry,  all with vanishing $\beta$ functions.\footnote{This statement is
slightly inaccurate; I do not want to dwell on subtleties.}

The next step was made   by Bershadsky et al. \cite{betal}.  These authors
eventually abandoned  AdS/CFT, and string methods at large.
Analyzing gauge field theories {\em per se} they proved
 that  an infinite set of amplitudes 
  in the orbifold daughters of the parent ${\cal N}=4$ theory
 in the large-$N$ limit coincide 
with those of the parent  theory, 
order by order in the gauge coupling. Thus, 
explicitly different theories have the same planar limit, at least perturbatively.

After a few years of relative oblivion, interest in the
issue of planar equivalence was revived by Strassler \cite{stras1}.
He shifted the emphasis away 
from the search   
for supersymmetric daughters, towards engineering QCD-like daughters.
 Strassler considered $Z_N$ orbifolds.
In 2003
an orientifold daughter of SUSY gluodynamics was suggested as
a prime candidate for nonperturbative equivalence \cite{asv1,asv}.
At $N=3$ this orientifold daughter identically reduces to
one-flavor QCD! Thus, one-flavor QCD is planar-equivalent to
SUSY gluodynamics. This remarkable circumstance allows one to copy
results of these theories from one to another.
For instance, color confinement of one-flavor QCD to supersymmetric Yang--Mills,
and the exact gluino condensate in the opposite direction. 
This is how the quark condensate was calculated,
for the first time analytically, in one-flavor QCD \cite{asv2}.

\subsection{Recent Developments}

Kovtun, \"{U}nsal and Yaffe formulated (and derived) \cite{kuy,kuy1}
the necessary and sufficient conditions
for nonperturbative planar equivalence to be valid.  This condition is nonbreaking of 
discrete symmetries: interchange $Z_2$ invariance
for the $Z_2$
orbifold daughter, and $C$ invariance 
for the orientifold daughter.
Although at first glance it does not seem to be a hard
problem to prove that  spontaneous breaking of the discrete symmetries
does not occur, 
in fact, this is a challenging problem which defies exhaustive 
solution so far. 

The question of the discrete symmetry nonbreaking
would be automatically solved if one could
prove that the expansion in fermion loops (say, for the vacuum energy) is
convergent in some sense \cite{asv3}.

To be more exact, let us
 give a mass term $m$ to the fermions, and assume at first this mass
term to be large, $m\gg\Lambda$.  Then the $N_f$ expansion is certainly convergent.
The question is ``is there a singularity, so that at small
$m$ the convergence is lost?"  I believe
that there is no such singularity. If so,
both $Z_2$ orbi and orienti-S/AS
are nonperturbatively equivalent to
supersymmetric gluodynamics. Note that this statement does not refer to $Z_N$ orbi
 with $N>2$.  In this case no mass term is
possible in the orbi theory, it is chiral.

On what I base my belief?
Consider supersymmetric gluodynamics with SUSY slightly broken by
a small mass term of gluino.  At $N=\infty$
the vacuum structure of this theory is exactly the same as the vacuum structure
of pure Yang--Mills (the latter was derived by Witten \cite{Witten:1998uk}, 
see also \cite{afterw,afterw1}).
Thus, I would say that the expansion in the number of fermion loops
should work.
This is of course not a mathematical theorem, but rather a physics argument.

Since for given number of fermion loops and given $m  \neq 0$
each expansion term in supersymmetric gluodynamics
is exactly the same as the corresponding expansion term
in $Z_2$ orbi and orienti-S/AS,
the fermion loop expansions in all three theories must be convergent.

Since in pure gauge theory, with no fermions,  the vacuum is 
unique \cite{Witten:1998uk},
then so is the case for $Z_2$ orbi and orienti-S/AS at $m \neq 0$.
The uniqueness of the vacuum state (for $\theta =0$)
implies the absence of the spontaneous breaking of the discrete symmetries
in the above daughter theories.

If the statement is valid for small $m \neq 0$
extrapolation to $m=0$ must be smooth
since none of these theories has massless particles in the
limit $m=0$. They all have a mass gap 
$\sim\Lambda$.

\subsection{Center-group symmetry and the limit $N\to\infty$}

The planar equivalence between the parent and daughter theories described
in the beginning of Sect.~\ref{pleq} holds not only on $R_4$
but in arbitrary geometry. Therefore, one can compare phase 
diagrams and, in particular, temperature dependences.
This topic was open by Sannino \cite{san}, a thorough discussion
was presented  by \"{U}nsal \cite{uns}.

There  is a famous Polyakov  criterion regarding
confinement/deconfinement in SU$(N)$ Yang--Mills theories.
If one compactifies $R_4$ into $R_3\times S_1$
and considers the Polyakov line along the compactified direction,
its expectation value may or may not vanish.
If it does not vanish, the $Z_N$ symmetry ---
the center of the gauge SU$(N)$ group ---
is broken. On the other hand, if the Polyakov line vanishes
the $Z_N$ symmetry is unbroken.
The former case corresponds to deconfinement, the latter to confinement.

Introducing quarks in the Yang--Mills Lagrangian
brings in a problem with this criterion,
since now there is no apparent $Z_N$ center symmetry at the Lagrangian level.
This is in one-to-one correspondence with the fact that
there are no genuine long strings in QCD. They break through 
quark-antiquark pair creation.

How can this apparent absence of the $Z_N$ symmetry be compatible
with planar equivalence?  In \cite{yaya} I argued that the center symmetry
is dynamically restored  in the
$N=\infty$ limit. In SU$(N)$ Yang--Mills theory with quarks
in the fundamental representation ($N\to\infty$)
the fundamental quarks decouple, and we are left with pure Yang--Mills
which does have the $Z_N$ center group.
Once it is unbroken, the theory is in the confinement phase. The Polyakov line is
a good order parameter. The same is valid with regards to supersymmetric gluodynamics
even at finite $N$. Gluinos do not decouple at large $N$,
but they do not ruin the center-group symmetry.

Now, comes a nontrivial remark.
Consider, for instance, the AS orientifold daughter
of supersymmetric gluodynamics.
At $N=\infty$ two-index antisymmetric fermions
do {\em not}  decouple. There is no apparent center 
symmetry in this theory.
At the Lagrangian level
the orientifold theories have at most a $Z_2$ center. 

And yet, the
full $Z_N$ center  symmetry dynamically emerges in the orientifold theories \cite{yaya2} in the limit
$N\to\infty$.
In the confining phase the manifestation of this enhancement is the existence
of stable $k$-strings  in the large-$N$ limit of the
orientifold theories. These strings are identical to those of supersymmetric 
Yang--Mills theories. The critical temperatures 
of the confinement-deconfinement phase transitions are the same in
the orientifold  daughters and their supersymmetric parent up to $1/N$ corrections.  

The Lagrangian of the orientifold theories has the form
\beq
{\mathcal L}= -\frac{1}{4g^2} \, G_{\mu\nu}^a G_{\mu\nu}^a
+\frac{i}{g^2}\, \psi_{ij}^{\alpha} D_{\alpha\dot\beta}\bar\psi^{\dot\beta\,\, ij}
\label{loor}
\eeq
where $\psi_{ij}$ is the Dirac spinor in the two-index
antisymmetric or symmetric representation. 
The center symmetry is $Z_2$ for even $N$ and none for odd $N$.  

 Integrating out the two-index antisymmetric fermion yields 
\begin{eqnarray} 
&& \log \det \left(i \gamma_{\mu} D_{\mu}^{\rm AS} - m\right) 
\nonumber\\[3mm]
&&
=  N^2 \,\sum_{n \in Z} \, \sum_{C_n} \,\frac{\alpha(C_n)}{2}\left\{
\left[\frac{\Tr}{N} U(C_n)\right]^2 -  \frac{1}{N} \frac{\Tr}{N} U^2(C_n)\right\}\,.
\label{Eq:detAS}
\end{eqnarray} 
In the large-$N$ limit we can ignore the single-trace 
terms since they are suppressed by $1/N$
compared to the $O(N^2)$ double-trace term.
The single-trace term contribution scales as that of the
fundamental fermions, and is quenched in the same manner. 
   
A typical double-trace  term $(\Tr U(C_n))^2$ is $O(N^2)$ and is a part of the leading large-$N$ dynamics.  Thus, the impact of the two-index antisymmetric fermions on dynamics is as important as that of  the glue sector of the theory. 

The action of the pure glue sector is local and manifestly invariant under the  $Z_N$ center.  Integrating out fermions, induces a nonlocal sum  (\ref{Eq:detAS}) over gluonic observables.  
This sum includes both topologically trivial loops with no net winding around the 
compact direction  (the $n=0$ term) and nontrivial loops with non-vanishing winding numbers.  
The topologically trivial loops are singlet under the $Z_N$ center symmetry by construction, while  the loops with non-vanishing windings are non-invariant. 

Let us inspect the  $N$ dependence of the effective action more carefully. 
If we expand the fermion action in the given gluon background  we get
\beq 
\left\langle \exp \left\{- N^2 
\sum_{n \neq 0} \,\sum_{C_n} \frac{\alpha(C_n)}{2}
\left[\frac{\Tr}{N} U(C_n)\right]^2   \right\}    \right\rangle\,,
\label{Eq:exp} 
\eeq
where $\langle ... \rangle$ means averaging
with the exponent combining the gluon Lagrangian with the zero winding number term. 
This weight function is obviously center-symmetric.
If  $h$ is an element of the SU$(N)$ center,  a typical term in the sum (\ref{Eq:exp}) 
transforms as 
\beqn
&& \left\langle \frac{\Tr}{N} U(C_n) \frac{\Tr}{N} U(C_n)\right\rangle  \longrightarrow h^{2n} \left\langle \frac{\Tr}{N}
 U(C_n) \frac{\Tr}{N}  U(C_n) \right\rangle \nonumber\\[4mm]
&&
=
h^{2n}  \left[ \left\langle \frac{\Tr}{N} U(C_n)\right\rangle \left\langle  \frac{\Tr}{N} U(C_n) \right\rangle +   \left\langle \frac{\Tr}{N} U(C_n) \frac{\Tr}{N} U(C_n) \right\rangle_{\rm con}  \right] ,\nonumber\\
 \label{Eq:factor}
 \eeqn 
 where I picked up a quadratic term as an example.
 The connected term in the expression above  is suppressed relative to the leading factorized part 
by  $1/N^2$, as follows from the standard $N$ counting, and can be neglected at large $N$.
As for the factorized part,
 planar equivalence implies that all expectation values 
of multi-winding Polyakov loops 
are suppressed in the large $N$ limit  by $1/N$, 
 \beqn 
&& \left\langle \frac{1}{N} \Tr U (C_n) \right\rangle^{\rm SYM} = 0\,,  
\nonumber\\[3mm]
&&  \left\langle \frac{\Tr}{N} U(C_n) \right\rangle^{\rm AS} = O\left(\frac{1}{N}\right)  \to 0\,,  \quad n \in Z- \{0\}\,,
\label{Eq:lowT}
\eeqn 
where the first relation follows  from unbroken  center symmetry in the SYM theory
and 
the latter is a result of planar equivalence ({\em in the $C$-unbroken,  confining phase of orienti-AS}). 

Consequently,  the  non-invariance of the expectation value of the action  under a global center transformation  is  
\beqn 
\langle \delta S \rangle = \langle S(h^n \Tr U(C_n))  -S ( \Tr U(C_n))  \rangle = O\left(\frac{1}{N}\right)  
\langle S \rangle \,,
\label{Eq:noninv}
\eeqn 
which implies, in turn, dynamical emergence  
of center symmetry in orientifold theories in
the large-$N$ limit. 
Let us emphasize again
that the fermion part of the Lagrangian
 which explicitly breaks the $Z_N$ symmetry  is {\em  not} sub-leading in large 
$N$.  However, the effect of the $Z_N$ breaking on  physical observables  is suppressed at
$N\to\infty$. 

This remarkable phenomenon is a natural (and straightforward) consequence of the large-$N$ equivalence  between ${\cal N} =1 $ SYM theory and orienti-AS.  Despite the fact that the center symmetry in the  orienti-AS Lagrangian is at most $Z_2$,  in the $N=\infty$ limit all
observables behave as if they are under the protection of the
$Z_N$ center symmetry. 
We refer to this  emergent symmetry of the orienti-AS vacuum  as  
 the   {\it custodial symmetry}.    
The custodial symmetry becomes   {\it exact}  in the  $N=\infty$ limit,
 and is {\it approximate} at large $N$.

Thus, we do have a $Z_N$ center-group symmetry in large-$N$ Yang--Mills theories
with quarks in one and two-index representations of SU$(N)$ after all!
Conceptually, this is a non-trivial statement which
 invalidates some statements in the literature; in particular,
it restores ``equal-rights" status for even and odd values of $N$.
 
\section {Exact perturbative calculations
with gluons in ${\mathcal N}=4$}

Obtaining  high orders in the perturbative expansion
(multiparton scattering amplitudes at tree level and with loops)
is an immense technical challenge.  Due to the gauge nature
of interactions, the final expressions for such amplitudes
are orders of magnitude simpler than intermediate expressions.

In the 1990's Bern, Dixon and Kosower pioneered applying string methods
to obtain loop amplitudes in  supersymmetric theories. The observed simplicity of these results (generalizing the elegant structure of the Parke--Taylor amplitude \cite{PT})
led to an even more powerful approach based on unitarity.
Their work resulted in an advanced helicity formalism exhibiting a feature 
of the amplitudes, not apparent from the Feynman rules, 
an astonishing simplicity.
In 2003 Witten uncovered \cite{Witten:2003nn} a hidden and elegant mathematical 
structure in terms of algebraic curves in terms of twistor 
variables
in gluon scattering amplitudes: he argued that the 
unexpected simplicity could be understood 
in terms of twistor string theory.

This observation created a diverse and thriving community of theorists advancing perturbative calculations at  tree level and beyond, as it became clear
that loop diagrams in gauge theories have their own hidden symmetry structure.
Most of these results do not directly rely on twistors and twistor string theory,
except for some crucial inspiration. So far, there is no good name for this subject.
Marcus Spradlin noted that an unusually large fraction
of contributors' names
start with the letter B\,.\footnote{E.g. Badger, Bedford, Berger, Bern, Bidder, 
Bjerrum-Bohr,
Brandhuber, Britto, Buchbinder, ... (Of course, one should not forget
about Cachazo, Dixon,  Feng, Forde, Khoze, Kosower, Roiban,  Spradlin, 
Svr\v{c}ek, Travaglini,  Vaman,  Volovich,  ...). This reminds me of a joke of a proof
given by a physicist that almost all numbers are prime:
one is prime, two is prime, three is prime, five is prime,
while four is an exception just supporting the general rule.
} Therefore, perhaps, we should call it $B$ theory, with B standing for beautiful,
much in the same way as M in $M$ theory stands for magic.
I could mention  a third reason for ``$B$~theory":
 Witten linked the scattering amplitudes to a
topological string known as the ``$B$~model."

$B$ theory revived, at a new level, many methods of the pre-QCD era,
when S-matrix ideas ruled the world. For instance, in a powerful paper due to Britto, Cachazo, Feng and Witten (BCFW) \cite{BCFW}, tree-level on-shell amplitudes were shown in a very simple and general way to obey recursion relations.
Their proof was based only on  Cauchy's theorem and
general (factorization) properties of tree-level scattering!
The BCFW recursion relations gave us a way to calculate
scattering amplitudes without using any gauge fixing or unphysical
intermediate states. 
   
Although the ultimate goal of the $B$ theory is calculating QCD amplitudes,
the concept design of various  ideas and methods is carried out in supersymmetric
theories, which provide an excellent testing ground.
Looking at super-Yang--Mills offers a lot of
insight into how one can  deal with the problems in QCD. 

Of all supersymmetric
theories probably the most remarkable is 
${\cal N}=4$ Yang--Mills. Its special status is due to the fact that (a) it is conformal, and (b) 
in the planar strong coupling limit it is dual to string theory
on AdS$_5\times {\rm S}^5$.

I would like to briefly
review a remarkable progress that has been achieved in understanding
the gluon ``scattering amplitudes" in this theory. 
(For more detailed reviews, with exhaustive lists of references, see e.g.
\cite{62}.) 
I refer to the scattering amplitudes in quotation marks because, strictly  speaking, the notion of the $S$ matrix
is ambiguous in conformal theories. This problem is resolved by regulating the
theory in the infrared. The standard regularization in this
range of questions is dimensional reduction to $D=4-2\epsilon$
dimensions.

\subsection{Bern--Dixon--Smirnov hypothesis}

In
2005 Bern, Dixon and Smirnov calculated,
in ${\cal N}=4$ theory, the
2 gluons $\to$ 2 gluons amplitude up to three loops
\cite{BDS}. Based on this and earlier results
with Anastasiou and Kosower \cite{anak}
they suggested an ansatz for the maximally helicity violating (MHV) $n$-point amplitudes to all 
loop orders in perturbation theory in the planar limit.
For 2 gluons $\to$ 2 gluons amplitude the Bern--Dixon--Smirnov (BDS) hypothesis takes the form
\beqn
&&{\cal A}(2\,\,{\rm gluons} \to 2\,\,{\rm gluons}) =
{\cal A}(2\,\,{\rm gluons} \to 2\,\,{\rm gluons})_{\rm tree}\times
\nonumber\\[3mm]
&&
\exp \left[ ({\rm IR\,\,\,divergent}) + \frac{\Gamma_{\rm cusp}(\lambda )}{4}\left(\ln\frac{s}{t}\right)^2
+{\rm const.} \right]
\label{bds}
\eeqn
where the infrared divergent $1/\epsilon^2,\,\,1/\epsilon$ part in the above expression is separated by virtue of the $\epsilon$ expansion, $\lambda$ is the 't Hooft coupling, and 
the function $f(\lambda )$ is directly related with the cusp anomalous dimension,\footnote{The cusp anomalous dimension was introduced in QCD in connection with
the renormalization-group equation for Wilson loops with cusps \cite{Polyakov},
see also \cite{Ivanov}.
It also controls the large-spin limit of the anomalous dimensions of twist-2
operators \cite{Ivanov,korc}.}
\beq
\Gamma_{\rm cusp} (\lambda ) =
\left\{
\begin{array}{l}
 \lambda -\frac{\pi^2}{12}\,\lambda^2 + ... ,\qquad \lambda\to 0\,,\\[3mm]
 {\sqrt\lambda}\,,\qquad \lambda\to \infty\,.
 \end{array}
 \right.
\eeq
Then  Alday and Maldacena
 \cite{AMal} in a  tour-de-force work 
performed  the {\em strong} coupling computation of the same amplitude by using the 
gauge theory/gravity duality that relates ${\cal N}=4$ Yang--Mills 
to string theory
on AdS$_5\times {\rm S}^5$. They found that the leading order result at large values of the 't Hooft coupling $\lambda$
is given by a single classical string configuration. The classical string solution 
corresponds to a minimal area $A_{\rm min}$ which, in turn, depends on the momenta $k_i^\mu$ of the final and initial gluons  $(i=1,2, ...,n)$,
\beq
{\mathcal A}_n/\left({\mathcal A}_{n}\right)_{\rm tree} \propto \exp\left\{-\sqrt{\lambda}
A_{\rm min}\right\}.
\eeq
The  Alday--Maldacena strong coupling result
for the finite part of the amplitude (\ref{bds}) turned out to be
exactly the same, $\exp\left\{(\Gamma_{\rm cusp}/2)\ln^2(s/t)\right\}$.
It should be noted that the cusp anomalous dimension  
$\Gamma_{\rm cusp}$ which fully determines the finite part of
the above amplitude is believed to be known exactly \cite{BES},
through a maximal transcedentality hypothesis.

The successful matching of the small and large-$\lambda$ expansions 
for the four-point amplitude was an
inspirational event, after which it was
natural to assume that the BDS representation holds
for arbitrary $n$-point amplitudes, $n=4,5,6,...$,
i.e.  the finite parts of all $n$-gluon amplitudes factorize into a $\lambda$ dependent
factor represented
by $\Gamma_{\rm cusp}$ and a momentum dependent factor, say, for $n=5$
\beq
{\mathcal A}_n/\left({\mathcal A}_{n}\right)_{\rm tree} \propto \exp\left\{
-\frac{\Gamma_{\rm cusp}(\lambda ) }{8}\,\sum_{i}
\ln\frac{s_{i,i+1}}{s_{i,i+2}}\,\ln \frac{s_{i+1, i+2}}{s_{i+2,i+3}} 
\right\},
\label{19}
\eeq
where $s_{l,l+1}$ are appropriate kinematic invariants.
Is this remarkably simple formula true?

Many people contributed to the solution of this question. Today
it is known  that the answer is negative, beginning from $n=6$,
and it is known why.  From the weak-coupling side,
the six-gluon amplitude calculated at two loops \cite{xx} reveals the occurrence of extra terms
absent in Eq.~(\ref{19}) which depend on conformally invariant ratios of various
$s_{i,j}$'s (see Sect.~\ref{mag}). 
Such ratios cannot be built for $n=4$ and 5.
From the strong-coupling side, the
classical string solution with appropriate boundary conditions
found  in the limit $n\to \infty$ \cite{AM3} also leads to a representation of the $n$-gluon amplitude incompatible 
with the BDS ansatz.

\subsection{A magical correspondence between the
coordinate and momentum spaces}
\label{mag}

The four-gluon amplitudes reveal an intriguing iterative structure both at weak
and strong couplings (i.e. in the
Bern--Dixon--Smirnov ansatz and the Alday--Maldacena approach, respectively). 
One may wonder where does this structure come from? 
The answer to this question was found in a series of papers by
Drummond, Henn, Korchemsky and Sokatchev (DHKS) \cite{D1,D2,D3,D4,D5} 
who found that the planar gluon
scattering amplitudes in ${\mathcal N}=4$ Yang--Mills possess a hidden symmetry, the so-called
{\em dual conformal symmetry}. This symmetry becomes manifest after one passes from on-shell gluon
momenta $p_i^\mu$ to dual four-dimensional ``coordinates" $x_i^\mu$,
\beq
p_i^\mu =x_i^\mu-x_{i+1}^\mu\,,\qquad \sum_i p_i^\mu=0\,,
\eeq
and considers the $n$-gluon scattering amplitude as a function of the dual coordinates $x_i$ 
(with the periodicity condition $x_{i+n}=x_i$). In this way, 
one discovers that, quite surprisingly, the
Feynman integrals contributing to four-gluon amplitudes up to four loops (!)
are invariant under the
conformal SO(2,4) transformations of the dual coordinates $x_i^\mu$.
If $x_i$ were ``normal"
coordinates in the configurational space, one could expect this symmetry to be related 
with the conformal symmetry of the underlying ${\mathcal N}=4$ theory. 
However, $x_i$'s belong to the momentum rather than configurational space!

In actuality, the dual conformal symmetry has a different
origin. Indeed, the dual coordinates are intrinsically related to momenta. Similar to conventional conformal symmetry, the dual conformal symmetry imposes severe constraints on the possible
form of the planar gluon amplitudes. Namely, 
as was shown by DHKS, if the dual conformal symmetry
survived to all loops, it would allow one to determine the four- and five-gluon amplitudes to all
orders. 

The dual conformality is slightly broken by infrared regularization and, as a
consequence, the scattering amplitudes satisfy anomalous conformal Ward identities. It is
remarkable that for the four- and five-gluon amplitudes the solution 
to these Ward identities is unique: 
it coincides with the Bern--Dixon--Smirnov ansatz. For 
$n\geq 6$ the Bern--Dixon--Smirnov ansatz for the MHV amplitudes goes through the same Ward
identities, 
but its general solution is determined up to an arbitrary function of (dual) conformal ratios.
If one assumes that the 
gluon scattering amplitudes enjoy the dual conformal symmetry to all orders, then
the failure of the BDS ansatz would imply that this function does not vanish.

\begin{figure}[h]
\begin{center}
\psfig{figure=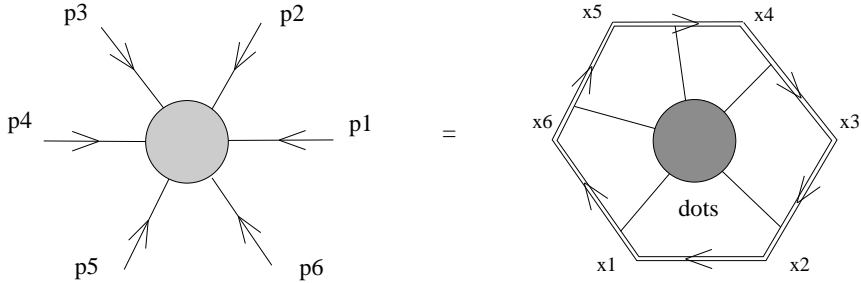,height=1.5in}
\end{center}
\caption{\small The $n$-gluon MHV amplitude $\langle 0| S| 1^- 2^-3^+\ldots n^+\rangle $
depicted on the left is equivalent to the Wilson loop $\langle 0|{\rm Tr}\, {\rm P} \exp \left( {i\oint_C dx\cdot
A(x)}\right)  |0\rangle $
with $n$ cusps
depicted on the right.}
\label{wl}
\end{figure}

Using the dual conformal symmetry as a guiding principle, DHKS suggested that the MHV
scattering amplitudes in planar ${\mathcal N}=4$ theory are equal to the Wilson loops evaluated
along a closed  polygon-like contour in the 
Minkowski space-time built from light-like segments defined
by on-shell gluon momenta (Fig.~\ref{wl}).

 This relation is very surprising and counter-intuitive because, firstly,
it relates two quantities of a different nature (an on-shell $S$-matrix element and the vacuum
expectation value of nonlocal functional of gauge fields) and, secondly, defining the integration
contour in terms of gluon momenta, one assigns a wrong engineering dimensions to
the  points in the
configuration space. Nevertheless, the two objects share the same symmetry (the dual conformal symmetry
of the scattering amplitudes versus conventional conformal symmetry of the Wilson loops in
${\mathcal N}=4$ theory). The duality  
between the $n=4$ and $n=5$ gluon amplitudes and the Wilson
loop is explicitly checked up to two loops in Refs.~\cite{D2,D3,B}. The
all-order proof presented in \cite{D3} is based on the conformal Ward identities. 

For $n\geq 6$ the
situation changes dramatically, since as was already mentioned, the conformal symmetry 
allows for the
existence of an arbitrary function of harmonic ratios. The dual conformal symmetry alone is not
sufficient to fix this function. It is natural to ask whether the duality with the Wilson loops will
still be valid.  

This question was addressed and clarified (for $n=6$ in two loops) in
Refs.~\cite{xx,D5}. We recall that the Bern--Dixon--Smirnov ansatz provides a definite prediction
for this amplitude. It was found that the BDS formula fails to describe the $n=6$
amplitude starting from two loops, and the discrepancy function  depends only on the (dual)
conformal ratios, in agreement with the dual conformal symmetry. 

Although the BDS fails for $n=6$, nevertheless one can ask about
 duality between the amplitude and the
light-like Wilson loop, as in Fig.~\ref{wl}. Comparing the two-loop expression for the hexagon Wilson loop with the
$n=6$ gluon amplitudes, one finds a perfect agreement! This strongly suggests that the Wilson
loops/scattering amplitudes duality should hold in ${\mathcal N}=4$ theory to all orders
and all values of $n$.

\section*{
Acknowledgments}

I am grateful to Adi Armoni, Zvi Bern, Sasha Gorsky, Gregory Korchemsky, 
David Tong, Mithat \"{U}nsal,
and  Alyosha Yung for valuable discussions. 
This work was supported in part by DOE grant DE-FG02-94ER408.

\end{document}